\documentclass[letterpaper, 10pt]{article} 
\usepackage{multicol}
\usepackage{lipsum}
\usepackage{subfig}
\usepackage[margin=2cm,a4paper]{geometry}
\usepackage{authblk}
\usepackage[utf8]{inputenc}
\usepackage[T1]{fontenc}
\usepackage{lmodern}
\usepackage{dcolumn}% Align table columns on decimal point
\usepackage{bm}% bold math
\usepackage{amsmath}
\usepackage{wrapfig}
\usepackage{graphicx}
\usepackage{verbatim}
\usepackage[UKenglish]{babel}
\usepackage{hyphenat}
\graphicspath{ {./pics/} }

\if TUTeX
  \usepackage{fontspec}
\else
  \usepackage[T1]{fontenc}
  \usepackage[utf8]{inputenc} % The default since 2018
  \DeclareUnicodeCharacter{200B}{{\hskip 0pt}}
\fi

\title{
Introdução à inferência Bayesiana: técnicas estatísticas para
análise de dados de íons pesados relativísticos
}
\author[1]{Liner Santos} 
\affil{
    Instituto de F\'{i}sica, Universidade de S\~{a}o Paulo, 
    R. do Mat\~{a}o 1371, 05508-090 S\~{a}o Paulo, SP, Brazil\\
    linersantos@usp.br\\
}
\author[2]{Thiago Domingues} 
\affil{
    Instituto de F\'{i}sica, Universidade de S\~{a}o Paulo, 
    R. do Mat\~{a}o 1371, 05508-090 S\~{a}o Paulo, SP, Brazil\\
    thiago.siqueira.domingues@usp.br\\
}
\date{}
\begin{document} 
\maketitle 
\begin{abstract}

Sob condições extremas de temperatura e pressão, acredita-se que os quarks e glúons (partículas mediadoras da interação entre quarks) podem estar ``livres'' em um determinado volume. Esta hipotética fase da matéria é chamada \textit{plasma de quarks e glúons}, QGP na sigla em inglês. Especula-se que tenha existido nos primeiros instantes após o \textit{Big Bang} e que exista no interior de estrelas de N\^eutrons devido \`a enorme densidade de energia nesses locais. Essas condições de altíssimas temperatura e densidade de energia podem ser reproduzidas em laboratório com a colisão de íons pesados em regime ultrarrelativístico em aceleradores como o RHIC e o LHC. Contudo, devido ao tempo extremamente curto de duração da fase QGP após a colisão, não conseguimos observar diretamente o plasma, apenas os chamados \textit{observáveis finais}, como as partículas geradas por esse conjunto de quarks, glúons e energia e a distribuição de momento dessas partículas. Assim, a modelagem matemática é uma ferramenta essencial no entendimento do comportamento do sistema, como por exemplo, para termos uma ideia do valor das viscosidades nesta fase.

\end{abstract}

\begin{abstract}

\emph{Under extreme conditions of temperature and pressure, it is believed that quarks and gluons (particles that mediate the interaction between quarks) can be ``free'' in a given volume. This hypothetical phase of matter is called \textit{plasma of quarks and gluons}, QGP for its acronym in English. It is speculated that it existed in the first moments after the \textit{Big Bang} and that it exists inside N\^eutron stars due to the enormous energy density in these places. These conditions of very high temperature and energy density can be reproduced in the laboratory with the collision of heavy ions in an ultra-relativistic regime in accelerators such as the RHIC and the LHC. However, due to the extremely short duration of the QGP phase after the collision, we were unable to directly observe the plasma, only the so-called \textit{final observables}, such as the particles generated by this set of quarks, gluons and energy and the distribution of momentum of these particles. Therefore, mathematical modeling is an essential tool in understanding the behavior of the system, for example, to have an idea of the value of viscosities in this phase.}
\end{abstract}

%\maketitle

\section{Introdução}

Conhecer quantitativamente e qualitativamente as propriedades das interações de quarks e glúons é uma tarefa extremamente complexa devido a pelo menos duas razões, a saber: 
Os métodos tradicionais empregados para realizar cálculos preditivos são os métodos de teoria perturbativa, que descrevem o espalhamento de duas partículas carregadas, que interagem fracamente, que não é o caso dos quarks, cuja interação é cerca de 100 vezes maior que a interação eletromagnética.

Não podemos realizar um experimento de espalhamento entre quarks e glúons, pois estes e aqueles não se propagam livremente no vácuo, estando sempre em estados ligados (confinados) dentro da matéria hadrônica (prótons, nêutrons, mésons, etc.). 

Assim, para investigarmos as propriedades dessas partículas devemos criar em laboratório condições extremas de temperatura e pressão que ``quebrem'' a estrutura hadrônica os quarks e gluons ``livres''. Hipoteticamente, essas condições estavam presentes nos primeiros estágios do nosso Universo e experimentos de colisões de íons pesados relativísticos podem, a princípio reproduzir essas condições extremas.

Aceleradores de partículas como o Relativistic Heavy Ion Collider (RHIC), Large Hadron Collider (LHC), que são fruto da colaboração de diversos países, realizam estes experimentos e nos fornecem dados para estes e outros estudos.

O progresso no conhecimento das colisões é devido à habilidade na construção de modelos fenomenológicos realísticos para simular o que ocorre durante a evolução temporal deste sistema. Os modelos físicos desenvolvidos para descrever o que ocorre durante os diversos estágios destas colisões são então comparados com os dados experimentais (os observáveis) produzidos nos aceleradores. Nós temos acesso apenas ao estágio final da colisão, no qual cerca de 10000 partículas chegam aos detectores em ângulos diferentes. A partir desse estado final de partículas produzidas durante a colisão, temos que inferir o que ocorreu em todo o processo físico.

Os modelos que descrevem a evolução temporal (linha do tempo) do meio
de íons pesados envolvem múltiplos estágios, em que cada estágio engloba
um grande conjunto de parâmetros incertos, descritos por uma dinâmica
potencialmente diferente. Esses tipos de modelos são chamados de modelos híbridos e envolvem 5 submodelos com um total de quase 20
parâmetros.

%\section{Modelo Híbrido}
%Como apontamos, a grande dificuldade \\
% $\bullet\bullet\bullet$
%\section{Inferência Bayesiana}

A inferência Bayesiana é uma metodologia estatística que pode nos fornecer uma maneira de lidar com problemas complexos de inferência. 
Suponha que temos um conjunto de observáveis experimentais, como $v_n$, que está relacionado à anisotropia na distribuição das partículas emergentes do processo de colisão, conforme explicado em \cite{Santos:2023rbef} e $p_T$, que é o momento das partículas emergentes na direção perpendicular ao eixo das velocidades relativas dos núcleos envolvidos na colisão. Estes são exemplos dos observáveis detectados ao fim de um evento de colisão entre dois íons pesados\cite{Santos:2023rbef} e este conjunto será representado por $D$ e representaremos por $\theta$ o conjunto de parâmetros que regem a evolução do sistema, como $\eta/s$, $\zeta/s$, etc.

Devido à impossibilidade de conhecimento do comportamento dos parâmetros englobados por $\theta$, faz-se uso da inferência Bayesiana, a partir de conhecimentos prévios, para a obtenção de uma distribuição de probabilidade desses parâmetros e se ter uma ideia sobre a evolução do sistema.

Este método indireto de obtenção dos parâmetros usa o teorema de Bayes
\begin{eqnarray}
    \mathcal{P}(\theta|D) = \frac{ \mathcal{P}(D|\theta) \mathcal{P}(\theta) }{ \mathcal{P}(D) },
\end{eqnarray}
em que

\begin{itemize}
    \item Distribuição a posteriori $\mathcal{P}(\theta|D)$: É a distribuição de probabilidade dos parâmetros que nos dão os valores detectados dos observáveis. Em outras palavras, qual o comportamento mais provável da viscosidade de cisalhamento $\eta/s$, dado os valores obtidos dos observáveis.
    \item Likelihood(ou função de verossimilhança) $\mathcal{P}(D|\theta)$: representa a probabilidade de observar algum valor dos dados experimentais dado um valor particular dos parâmetros do modelo.
    \item Distribuição a priori $\mathcal{P}(\theta)$: Representa nosso conhecimento prévio sobre os parâmetros englobados por $\theta$. 
    \item Evidência ($\mathcal{P}(D)$): Um fator de normalização. 
\end{itemize}
Como estamos interessados apenas na inferência da distribuição de valores assumida por nosso parâmetro, podemos usar a seguinte relação
\begin{eqnarray}
\mathcal{P}(\theta|D) \propto \mathcal{P}(D|\theta) \mathcal{P}(\theta),
\label{propto}
\end{eqnarray}
que não leva em conta a normalização.
Podemos aplicar esses conceitos utilizando um exemplo relacionado ao estudo de colisões de íons pesados relativísticos. 
No estudo destas colisões, os observáveis chamados coeficientes de fluxo $v_n$\cite{Voloshin:1994mz} são sensíveis às viscosidades do Plasma de Quarks e Glúons(QGP)\footnote{É justo salientar que o plasma é um fluido e sua evolução é regida pela hidrodinâmica relativística\cite{Romatschke:2017ejr} e as viscosidades são propriedades dos fluidos.}. Em outras palavras, podemos parametrizar os coeficientes de fluxo em termos das viscosidades, além dos parâmetros que definem as condições iniciais, lembrando que há uma relação entre esses coeficientes e a anisotropia inicial do sistema\cite{Santos:2023rbef}.

\subsection{Definindo um modelo físico simples}
Para aplicar os conceitos da inferência Bayesiana, começaremos definindo um modelo físico simples. Vimos que a função verossimilhança $\mathcal{P}(\theta|D)$ codifica a probabilidade condicional de observar algum valor dos observáveis finais, como os coeficientes de fluxo por exemplo, dado um valor particular dos parâmetros do modelo, como as viscosidades\footnote{A rigor, o parâmetro que analisamos nas colisões de íons pesados relativísticos são as razões entre as viscosidades e a entropia\cite{Ollitrault:2007du}.} $\frac{\eta}{s}$ e $\frac{\zeta}{s}$. 

Vamos utilizar uma aproximação linear para o nosso modelo físico que nos permita focar na discussão de conceitos da inferência Bayesiana e emulação. No entanto, devemos ter em mente que um modelo físico mais realista pode ser arbitrariamente complexo, não linear e demandar muito esforço computacional.  

Além da linearidade entre os parâmetros e os observáveis, é conveniente adicionar um erro estatístico, sem correlações, em cada predição do fluxo elíptico para entendermos como erros estatísticos influenciam nosso problema de inferência.

Como um exemplo, vejamos o caso em que queremos obter apenas o fluxo elíptico e iremos supor que este dependa apenas da viscosidade de cisalhamento e esta não varia ao longo da evolução do sistema. Neste caso, seja $y$ a saída do nosso modelo ($v_2$) e $\theta$ a variável independente, isto é, os valores assumidos pela viscosidade de cisalhamento específica. Podemos escrever então, 
%Como um exemplo, vejamos um modelo simplificado em que é usado o ajuste linear entre o parâmetro e os observáveis
%
\begin{eqnarray}
    y = m \cdot \theta + b +\epsilon, \nonumber\\
    \Rightarrow v_2 = m\cdot(\frac{\eta}{s}) + b + \epsilon
    \label{eq1}
\end{eqnarray}
onde $m$, $b$ e $\epsilon$ são coeficientes reais e o erro estatístico, respectivamente. 

Conforme dito, para termos resultados razoáveis, este cálculo deveria ser feito muitas vezes $(~\mathcal{O}(10^6))$ e isso torna o problema computacionalmente muito custoso e demorado,  uma vez que os resultados dessa expressão são obtidos mediante a simulação completa de uma colisão. Assim, para contornar este inconveniente, faz-se uso de emuladores, que nos retornam um mapeamento da equação (\ref{eq1}), a partir de poucos cálculos diretos. 

A seguir, é explicado como é feita essa escolha dos valores a serem calculados diretamente e como é feito o mapeamento por meio do emulador.

\begin{comment}
Modelos físicos possuem incertezas, o poder de conseguir quantificar incertezas é uma das principais qualidades da abordagem Bayesiana. Podemos adicionar um erro estatístico sem correlações para cada variável de saída da nossa simulação. Sendo assim, denotando a saída do modelo por $y$ e $\theta$ o parâmetro de interesse, o nosso modelo linear pode ser escrito como $$y = m \cdot \theta + b +\epsilon,$$ 
onde $m$, $b$ e $\epsilon$ são as variáveis independentes e o erro estatístico, respectivamente. 
\end{comment}

%$\bullet\bullet\bullet$

\subsection{Pontos de Design}

Em colisões de íons pesados relativísticos, utilizamos hidrodinâmica viscosa como um modelo físico para evoluir o perfil inicial de densidade de energia. Os códigos numéricos podem levar horas para simular um único evento
para algum ponto no espaço de parâmetros. Além disso, temos que simular
um grande conjunto de eventos para calcular os observáveis com precisão
estatística suficiente. No entanto, quando estamos lidando com modelos
que demandam um grande poder computacional, podemos empregar um
substituto para este modelo que consiga estimar as suas saídas de uma
maneira mais rápida e fornecendo a incerteza deste ``modelo substituto''.

Quando estamos lidando com modelos que demandam um grande poder computacional, podemos empregar um substituto para este modelo que consiga estimar as suas saídas de uma maneira mais rápida e fornecendo a incerteza deste "modelo substituto". Essa técnica é empregada utilizando processos Gaussianos, ou seja, é uma interpolação não paramétrica.
Como em toda interpolação, precisamos de um conjunto de pontos (uma
amostra) no espaço de parâmetros, tal que seja possível conhecer a saída
(os resultados) do nosso modelo físico. Sendo assim, precisamos utilizar
nosso modelo físico simples para calcular as saídas no conjunto de pontos
do nosso espaço de parâmetros. Vamos chamar essa amostra de nossos \textbf{pontos de Design}.

%%%% ----- Figura mostrando cálculos de fluxo elíptico --- %%%%
%
%\begin{figure}[!htb]
%    \centering
%    \includegraphics[width=\linewidth]{pictures/latin.png}
%    \caption{Saídas da simulação numérica utilizando nosso modelo simplificado. Calculamos 30 pontos de design  com 10 % de erro estatístico}}
%    \label{Model Design Predictions}
%\end{figure}
%

\begin{figure}[!h]
    \centering
    \includegraphics[width=10.0cm]{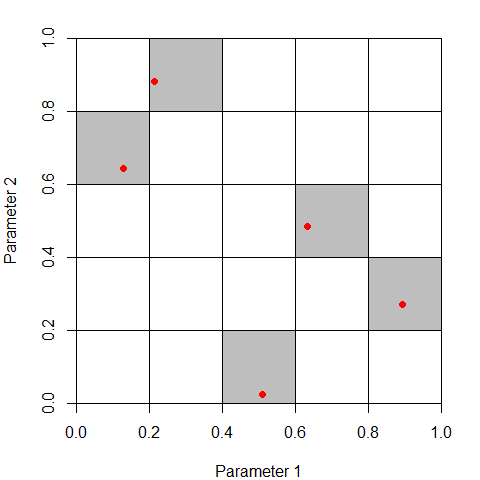}
    \caption{Ilustração do uso da técnica do hipercubo latino. Imagem obtida de \textit{cran.r-project.org}}
    \label{hiperlat}
\end{figure}

Para a escolha dos pontos de design, geralmente é usada a técnica do hipercubo latino, em que cada cada linha e coluna contém apenas um ponto. Em outras palavras, o espaço de parâmetros é dividido em N linhas e N colunas de modo que não há dois pontos em uma mesma linha ou coluna\footnote{Esta divisão em linhas e colunas tem apenas fins didáticos pois em geral estamos num espaço com n dimensões.}. A fig.(\ref{hiperlat}) ilustra esta técnica para o caso bidimensional.

Agora, precisamos interpolar entre os pontos de design para obter saídas para todo o espaço de parâmetros (número infinito de pontos).

\subsection{Emulador Gaussiano: um proxy para modelos computacionais custosos}

No estudo das colisões de íons pesados, os observáveis, de acordo com o modelo hidrodinâmico\cite{Ollitrault:2007du}\cite{DerradideSouza:2015kpt}, dependem de muitos parâmetros\footnote{17 parâmetros de acordo com o modelo híbrido da JETSCAPE\cite{JETSCAPE:2020mzn}.} e o mapeamento dessa relação, i.e. encontrar a relação entre os observáveis e os parâmetros não é uma tarefa simples. Em outras palavras, um ajuste de reta N-dimensional não é uma boa escolha. 

Além disso, devemos ter muitos valores dos observáveis, com os respectivos parâmetros. Isso tornaria o mapeamento uma tarefa impossível, até para os computadores mais potentes. Para contornar este problema, fazemos uso dos emuladores, que são processos que `imitam' o comportamento do modelo com poucos valores de entrada e com razoável precisão, embora isso requeira uma escolha correta da função que será usada no processo de emulação. O processo de emulação mais usado é o chamado \textit{processo Gaussiano}. Este processo tem a propriedade de que qualquer ponto aleatório obedece à distribuição normal.
\begin{eqnarray}
y \sim \mathcal{N}(\mu,\sigma)
\end{eqnarray}

Além disso, qualquer conjunto finito de seus pontos obedece a uma distribuição normal multivariada.
\begin{eqnarray}
\mathbf{y} \sim \mathcal{N}(\mbox{\boldmath{$\mu$}},\Sigma)
\end{eqnarray}
Nesta equação, \mbox{\boldmath{$\mu$}} representa o vetor médio e $\Sigma$ é a matriz de covariância, que mede as correlações entre pontos diferentes do espaço.
$$\Sigma = \left[
\begin{array}{c c c}
\sigma^2_{11}&\ldots& Cov(\textbf{x}_{1},\textbf{x}_n)\\
Cov(\textbf{x}_{2},\textbf{x}_1)& \ldots& Cov(\textbf{x}_{2},\textbf{x}_n)\\
\vdots&\ddots &\vdots\\ Cov(\textbf{x}_{n},\textbf{x}_1)&\ldots& \sigma^2_{nn}
\end{array}\right]
$$

A função usada no processo gaussiano, i.e. função que interpolará os pontos calculados (chamados \textit{design points}), tem a forma a seguir:

\begin{eqnarray}
exp[(x-m)^T k(\textbf{x,y})(\textbf{y-m})]
\end{eqnarray}

Nesta expressão, m é o valor médio de cada parâmetro $x_i$ e o termo $k(x,y)$ é o chamado \textit{núcleo} (ou \textit{kernel}) do emulador e é quem determina a covariância dos valores produzidos pelo emulador. \footnote{Para mais detalhes sobre os núcleos mais populares utilizados em Processos Gaussianos, confira \textit{https://www.cs.toronto.edu/~duvenaud/cookbook/} e \textit{https://distill.pub/2019/visual-exploration-gaussian-processes/} para explorar visualmente} A escolha do kernel depende de algum conhecimento prévio do comportamento esperado do sistema e pode ser trocado no caso dos resultados não serem satisfatórios. 

Devemos fornecer também algum intervalo de valores para os parâmetros do emulador, os chamados \textbf{
hiperparâmetros}. Quando ajustamos o emulador, os valores assumidos pelos hiperparâmetros serão obtidos maximizando a seguinte função verosimilhança:
$$\log p(y^*|y_{t}, \theta) \propto -\frac{1}{2}y_{t}^{T} \Sigma^{-1}_{y_t} y_{t} - \frac{1}{2} \log |\Sigma_{y_t}|$$
onde $\Sigma_{y_t}$ é a matriz covariância resultante de aplicarmos a função de covariância aos \textbf{dados de treinamento}. O primeiro termo desta função verossimilhança é responsável por `recompensar' o emulador com hiperparâmetros que se ajustam bem aos dados, enquanto o segundo termo penaliza um emulador que faz um sobreajuste (overfitting) nos dados.
%%% Precisamos explicar melhor o parágrafo acima: O que sao hiperarametros?, O que é este ajuste do emulador

Há algumas funções comumente utilizadas como núcleo e nós vamos utilizar aqui uma combinação entre a \textbf{função de base radial} (RBF)\footnote{Do Inglês \textit{radial basis function}.} e o \textbf{núcleo de ruído branco} (ou \textit{White Noise Kernel}). Este segundo núcleo é necessário porque um modelo físico possui acurácia estatística finita, isto é, há uma incerteza associada a cada ponto de design escolhido e o segundo núcleo engloba essas incertezas. Assim, o núcleo final é formado pela soma dos dois núcleos citados. \footnote{Existem outras combinações possíveis, poderíamos ter feito o produto ou outras composições das duas funções.}. 
%%% devemos complementar a expressão abaixo
\begin{eqnarray}
k(x,y) = exp\left(-\frac{|x - y|^2}{2l^2}\right)
\end{eqnarray}
com $l$ sendo o chamado \textit{comprimento de escala} e uma  

\begin{figure}[!h]
    \centering
    \includegraphics[width=10.0cm]{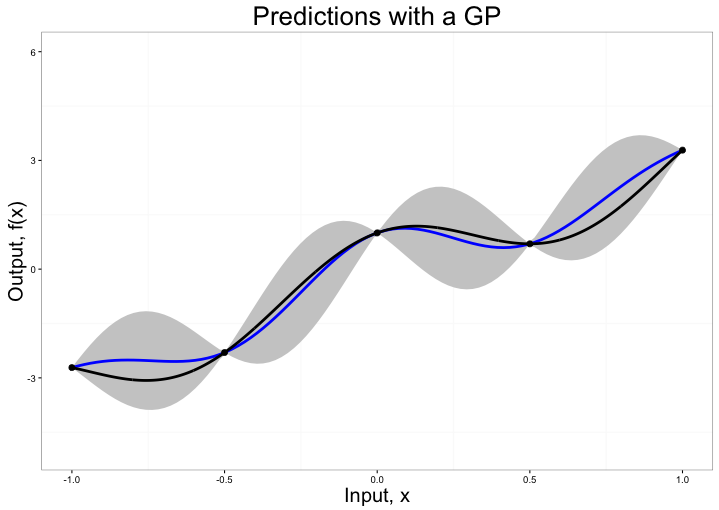}
    \caption{Exemplo de aplicação do processo Gaussiano para interpolar alguns pontos de design. A linha preta é a função escolhida como teste e a linha azul é formada pelos resultados fornecidos pelo emulador.}
    \label{gauemu}
\end{figure}

Em geral, é feito um escalonamento das variáveis de modo a termos média nula para simplificar a equação acima. A fig.(\ref{gauemu}) ilustra o uso dos emuladores e sua acurácia. Neste caso, sabemos qual a função verdadeira (linha preta) e escolhemos alguns pontos desta, que são os pontos de design. A partir destes pontos, o emulador calcula os valores intermediários, que estão indicados pela linha azul. A região sombreada é a incerteza sobre os valores calculados pelo emulador e isso é outra vantagem dos emuladores: eles nos informam também a incerteza sobre os valores calculados.

Após construirmos e treinarmos o emulador, precisamos checar quão bem o emulador consegue fitar o modelo físico. É esperado que o emulador consiga fitar os pontos para os quais ele foi treinado, mas queremos verificar se ele consegue prever pontos, no espaço de parâmetros, para os quais ele não foi treinado. Sendo, assim, precisamos realizar algumas validações do emulador, vamos utilizar um conjunto de pontos que vamos chamar de \textit{\textbf{conjunto de dados de teste}}. 

%#Plot the emulator prediction vs the hydro model prediction
%\begin{figure}
%    \centering
%    \includegraphics{}
%    \caption{Caption}
%    \label{fig:enter-label}
%\end{figure}

"Além disso, podemos checar se as previsões do emulador estão
enviesadas. Seja $\hat{y}(\theta)$ a previsão que o emulador fornece para os parâmetros
$\theta$ e $y(\theta)$ os cálculos numéricos do nosso modelo, então podemos definir
uma quantidade chamada resíduo $\hat{y}(\theta) - y(\theta)$ que se apresenta como uma
função dos parâmetros."

% plot the residual as a function of 
%\begin{figure}
%    \centering
%    \includegraphics{}
%    \caption{Caption}
%    \label{residual}
%\end{figure}
Para verificar a performance do emulador localmente no espaço de parâmetros, podemos visualizar a distribuição dos resíduos normalizada pelo incerteza do emulador $r \equiv (\hat{y} - y)/\hat{\sigma}$, onde $\hat{y}$ são os resultados do emulador em um ponto de validação, $y$ é o valor verdadeiro do modelo e $\hat{\sigma}$ é a incerteza do emulador neste mesmo ponto no espaço dos parâmetros. \footnote{Existem outros testes que podemos explorar para verificar a performance dos emuladores.} %\url{https://en.wikipedia.org/wiki/Q%E2%80%93Q_plot} para mais detalhes}   

\section{Inferência Bayesiana}

Depois que treinamos e validamos nosso emulador, temos uma maneira rápida e acurada de comparar com os dados experimentais em qualquer região do espaço de parâmetros. Agora, podemos utilizar o nosso emulador para performar uma inferência Bayesiana. Primeiramente precisamos definir nossa crença sobre os valores dos parâmetros antes de utilizar os dados experimentais.    

\subsection{Distribuição à Priori}
A distribuição a priori engloba o conhecimento prévio que se tem a respeito
dos parâmetros do modelo. Este conhecimento prévio, bem como a
determinação dos valores máximo e mínimo de cada parâmetro, pode provir
de vínculos teóricos, como por exemplo a positividade das viscosidades de
modo a não violar a Segunda Lei da Termodinâmica.\footnote{Viscosidade negativa implicaria em diminuição da entropia do sistema.}.

A escolha mais trivial é a distribuição uniforme que, muitas vezes, é a única possível por não termos acesso direto ao sistema\cite{JETSCAPE:2020mzn}. Supondo que todos os parâmetros são independentes, a distribuição de probabilidade a priori conjunta (\textit{joint prior distribution}) é dada por:
\begin{eqnarray}
    \mathcal{P}(\textbf{x}) = \prod_{i=1}^{N} p(x_i - x_{i_{min}})p(x_{i_{max}} - x_i)
\end{eqnarray}
sendo $x_i$ o i-ésimo parâmetro do modelo.

% Plot two different priors
%\begin{figure}
%    \centering
%    \includegraphics{}
%    \caption{Duas distribuições à priori.}
%    \label{prior distribution}
%\end{figure}
%
Para compararmos os cálculos numéricos do modelo com os dados experimentais, precisamos definir uma distribuição verossimilhança. 

\subsection{Distribuição Verossimilhança}
Distribuição de probabilidade que informa qual a probabilidade de obtermos um valor do observável, dado que os parâmetros têm um valor conhecido. É comum assumirmos que os erros experimentais se distribuam de uma maneira Gaussiana, ou seja, sigam uma distribuição Normal multivariada. Nas análises Bayesianas recentes em colisões de íons pesados, em geral se usa a distribuição normal multivariada para a verossimilhança\cite{JETSCAPE:2020mzn}:

\begin{eqnarray}
    \mathcal{P}(D|\theta) = [(2\pi)^n det(\Sigma)]^{-1/2}\cdot exp[-\frac{1}{2}\Delta y^T\Sigma^{-1}\Delta y].
    \label{multi}
\end{eqnarray}

Nesta equação, $\Sigma$ é a matriz de covariância dos observáveis, $n$ é a quantidade de observáveis (i.e., a dimensão do vetor $D$) e $\Delta y = y_{obs} - y_{modelo}$ é a diferença entre o valores calculados pelo modelo e os valores experimentais.

Em geral, é conveniente expressarmos o logaritmo da verossimilhança\cite{Nijs:2020roc}

\begin{eqnarray}
    ln[\mathcal{P}(D|\theta)] = -\frac{1}{2}ln[(2\pi)^n det(\Sigma)] -\frac{1}{2}\Delta y^T\Sigma^{-1}\Delta y
\end{eqnarray}

\subsection{Distribuição à Posteriori}
A distribuição a posteriori é o produto entre a distribuição a priori e a
função verossimilhança. Sendo assim, é mais conveniente tomar o
logaritmo natural desta distribuição, pois resultará na soma dos logaritmos
de ambas as distribuições.

Conforme mostrado na eq.(\ref{propto}), a distribuição a posteriori é determinada, a menos de um fator de normalização, pelo produto da distribuição a priori e da verossimilhança. Aqui também é conveniente expressarmos o logaritmo desta probabilidade

\begin{eqnarray}
    ln[\mathcal{P}(\theta|D)] = ln[\mathcal{P}(\theta)]+ln[\mathcal{P}(D|\theta)]
    \label{posterior}
\end{eqnarray}

No caso das colisões de íons pesados, a expressão (\ref{posterior}) pode ser escrita da seguinte forma:
\begin{eqnarray}
  ln[\mathcal{P}(\frac{\eta}{s},\frac{\zeta}{s}|v_2,v_3,v_4,v_5)] = \nonumber\\
  ln[\mathcal{P}(\frac{\eta}{s},\frac{\zeta}{s})]+ln[\mathcal{P}(v_2,v_3,v_4,v_5|\frac{\eta}{s},\frac{\zeta}{s})] 
\end{eqnarray}

Nesta expressão, consideramos que as viscosidades de cisalhamento $(\frac{\eta}{s})$ e volumétrica $(\frac{\eta}{s})$ são constantes ao longo da evolução do sistema\footnote{Em geral, adota-se um modelo em que há uma dependência da temperatura sobre as viscosidades e isso aumenta consideravelmente o número de parâmetros\cite{JETSCAPE:2020mzn}.}.

\subsubsection{Distribuição marginal}

Na análise do comportamento dos parâmetros que regem a evolução do sistema, a distribuição marginal desempenha um papel importante, uma vez que ela nos mostra os valores mais prováveis para cada parâmetro, e também nos dá uma ideia das correlações entre parâmetros.

Suponha que $\theta$ seja um vetor composto por $n$ parâmetros. A distribuição marginal de $\theta_1$, é definida por

\begin{eqnarray}
    \mathcal{P}(\theta_1|D) = \int_{\theta_2}...\int_{\theta_n} \mathcal{P}(\theta|D)
\end{eqnarray}

Da mesma forma, podemos calcular a distribuição marginal sobre dois ou mais parâmetros. Por exemplo, sobre $\theta_1$ e $\theta_2$:

\begin{eqnarray}
    \mathcal{P}(\theta_1,\theta_2|D) = \int_{\theta_3}...\int_{\theta_n} \mathcal{P}(\theta|D)
\end{eqnarray}

\begin{figure}[!h]
    \centering
    \includegraphics[width=10.0cm]{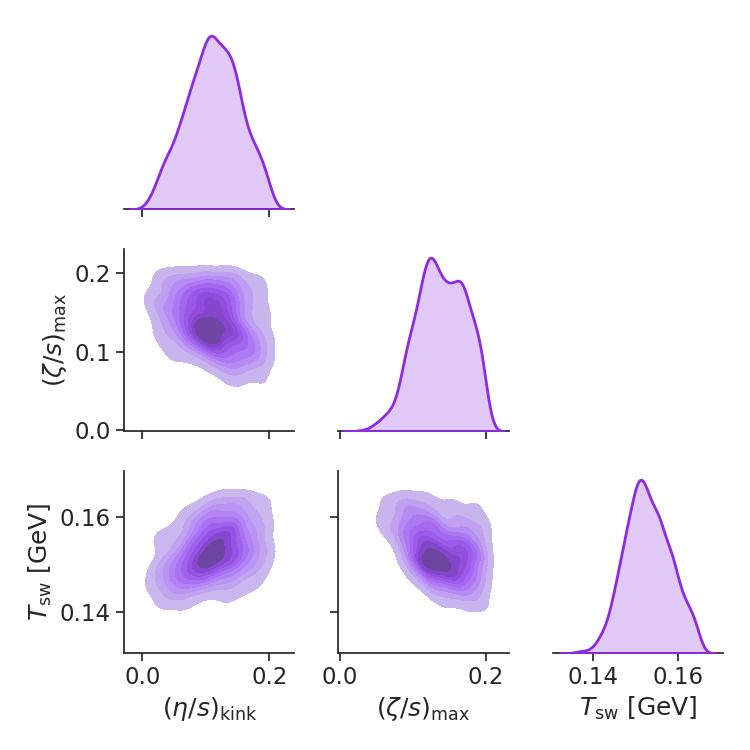}
    \caption{Distribuição marginal de alguns parâmetros usados no estudo da evolução do sistema.}
    \label{fig_posterior}
\end{figure}

A figura (\ref{fig_posterior}) mostra a distribuição marginal para alguns parâmetros que determinam os harmônicos de fluxo $v_n$,  de acordo com o modelo proposto em \cite{JETSCAPE:2020mzn}, em que as viscosidades variam com a temperatura. Vemos também a distribuição marginal entre os pares de parâmetros.

\section{Aplicações em colisões de íons pesados relativísticos}

\begin{figure}[!h]
    \centering
    \includegraphics[width=10.0cm]{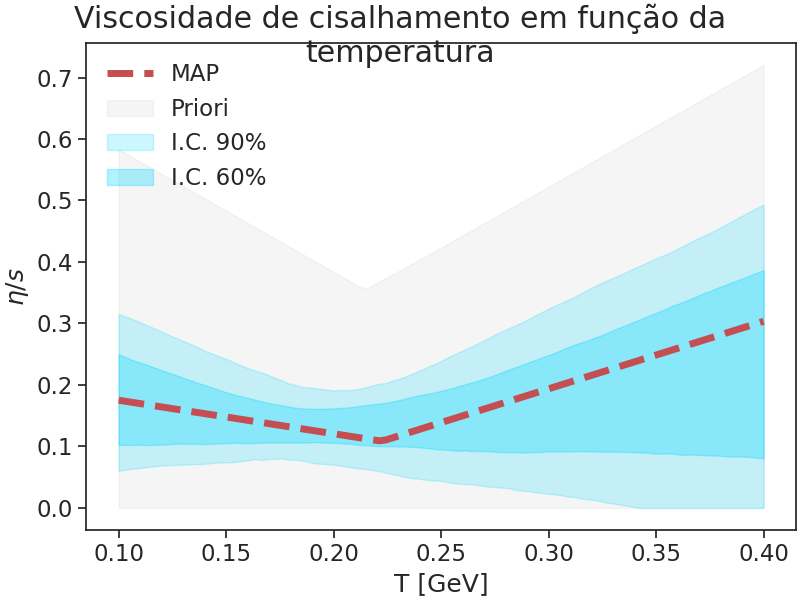}
    \caption{Viscosidade de cisalhamento e os intervalos de confiança. A região cinza indica a viscosidade calculada com os parâmetros a priori.}
    \label{eta-posterior}
\end{figure}

\begin{figure}[!h]
    \centering
    \includegraphics[width=10.0cm]{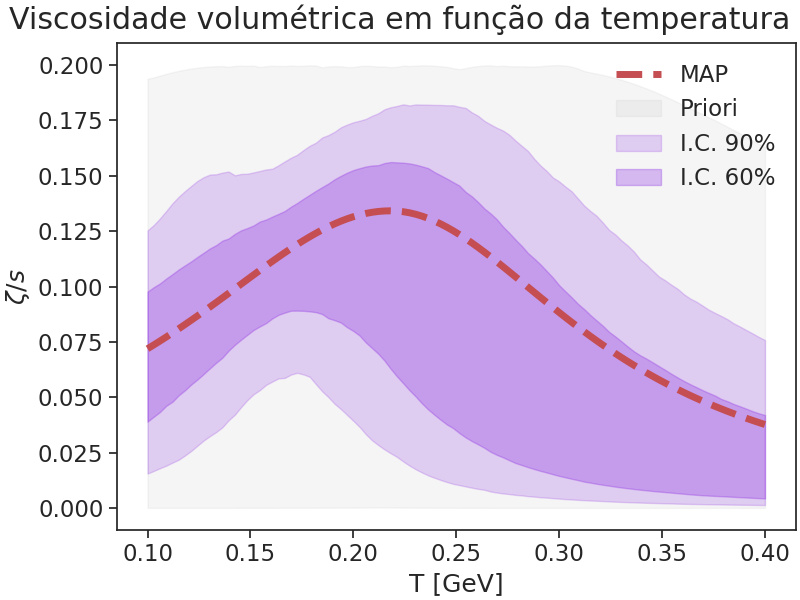}
    \caption{Viscosidade volumétrica e os intervalos de confiança. A região cinza indica a viscosidade calculada com os parâmetros a priori.}
    \label{zeta-posterior}
\end{figure}

As figuras (\ref{eta-posterior}) e (\ref{zeta-posterior}) mostram a distribuição de probabilidade para as viscosidades $\frac{\eta}{s}$ e $\frac{\zeta}{s}$ em alguns intervalos de confiança, I.C. Por exemplo, a região indicada por $90\%$, mostra que há $90\%$ de chance que o valor da viscosidade esteja nesta região para cada valor da temperatura, indicada no eixo horizontal. Vemos também os valores calculados com os parâmetros que dão o máximo a posteriori (MAP) para as viscosidades.

\section{Considerações finais}

Conforme mostrado em \cite{Santos:2023rbef}, o estudo das colisões ultrarrelativísticas de íons pesados nos mostra a relevância da inferência sobre um sistema que não conseguimos observar diretamente, a partir
dos observáveis medidos. Em outras palavras, usamos a inferência estatística, em particular o teorema de Bayes, para resolver o chamado ``problema inverso'' em que se tem os resultados finais e o objetivo é determinar o comportamento dos parâmetros do modelo usado para descrever a evolução do sistema. 

A inferência bayesiana nos permite obter o conhecimento prévio dos
parâmetros a partir de outros experimentos e de resultados teóricos
provenientes de diversos estudos. Essa característica se apresenta como
uma vantagem em relação à inferência frequentista, a qual não leva em consideração conhecimentos prévios acerca dos parâmetros envolvidos na evolução do sistema mas apenas em resultados de experimentos semelhantes.

O site \textit{https://github.com/JETSCAPE/SummerSchool2021} fornece vários códigos, na linguagem Python, com exemplos do uso da inferência bayesiana no estudo das colisões de íons pesados.

\section*{Agradecimentos}

Agradecemos aos professores André Giannini e Matthew Luzum e aos colegas Jefferson Arthur e João Paulo Pichetti pelas valiosas discussões.

%\printbibliography
\bibliographystyle{acm} \bibliography{referencias}

\begin{thebibliography}{1}

\bibitem{DerradideSouza:2015kpt}
{\sc Derradi~de Souza, R., Koide, T., and Kodama, T.}
\newblock {Hydrodynamic Approaches in Relativistic Heavy Ion Reactions}.
\newblock {\em Prog. Part. Nucl. Phys. 86\/} (2016), 35--85.

\bibitem{JETSCAPE:2020mzn}
{\sc Everett, D., et~al.}
\newblock {Multisystem Bayesian constraints on the transport coefficients of
  QCD matter}.
\newblock {\em Phys. Rev. C 103}, 5 (2021), 054904.

\bibitem{Nijs:2020roc}
{\sc Nijs, G., van~der Schee, W., G\"ursoy, U., and Snellings, R.}
\newblock {Bayesian analysis of heavy ion collisions with the heavy ion
  computational framework Trajectum}.
\newblock {\em Phys. Rev. C 103}, 5 (2021), 054909.

\bibitem{Ollitrault:2007du}
{\sc Ollitrault, J.-Y.}
\newblock {Relativistic hydrodynamics for heavy-ion collisions}.
\newblock {\em Eur. J. Phys. 29\/} (2008), 275--302.

\bibitem{Romatschke:2017ejr}
{\sc Romatschke, P., and Romatschke, U.}
\newblock {\em {Relativistic Fluid Dynamics In and Out of Equilibrium}}.
\newblock Cambridge Monographs on Mathematical Physics. Cambridge University
  Press, 5 2019.

\bibitem{Santos:2023rbef}
{\sc Santos, L.}
\newblock {Colis\~oes de \'\i{}ons pesados \textendash{} Um exemplo de
  observa\c{c}\~ao indireta de fen\^omenos f\'\i{}sicos}.
\newblock {\em Rev. Bras. Ens. Fis. 45\/} (2023), e20230127.

\bibitem{Voloshin:1994mz}
{\sc Voloshin, S., and Zhang, Y.}
\newblock {Flow study in relativistic nuclear collisions by Fourier expansion
  of Azimuthal particle distributions}.
\newblock {\em Z. Phys. C 70\/} (1996), 665--672.

\end{thebibliography}

\end{document}